\documentclass[journal]{IEEEtran}

\usepackage{xcolor,soul,framed} 

\colorlet{shadecolor}{yellow}
\usepackage[cmex10]{amsmath}
\usepackage{amssymb}
\usepackage{array}
\usepackage{mdwmath}
\usepackage{mdwtab}
\usepackage{eqparbox}

\usepackage{graphicx}
\graphicspath{{../pdf/}{../jpeg/}}
\DeclareGraphicsExtensions{.pdf,.jpeg,.png}
\usepackage{caption} 
\usepackage{subcaption}

\usepackage{algorithm}
\usepackage{algorithmic}

\usepackage{url}

\usepackage[pdfencoding=auto]{hyperref}
\hypersetup{
	colorlinks=true,
	linkcolor=blue,
	anchorcolor=blue,
	citecolor=blue
}

\begin{document}
	\bstctlcite{IEEEexample:BSTcontrol}
	\title{Enhancing User Fairness in Two-Layer RSMA: A Movable Antenna Approach}
	
	\author{Ji Luo, Yaxuan Chen, Guangchi Zhang, Miao Cui, Hao Fu, and Changsheng You
		\thanks{
			The work of G. Zhang and M. Cui was supported by Guangdong Basic and Applied Basic Research Foundation under Grant 2026A1515011208. The work of C. You was supported by the National Natural Science Foundation of China under Grant 62571227. (Corresponding author: Guangchi Zhang)
			
			Ji Luo, Yaxuan Chen, Guangchi Zhang, Miao Cui, Hao Fu are with the School of
			Information Engineering, Guangdong University of Technology, Guangzhou 
			510006, China (e-mail: \{luoji2, chenyaxuan\}@mails.gdut.edu.cn; \{gczhang, cuimiao, ecejasonhaofu\}@gdut.edu.cn). Changsheng You is with the Department of Electronic and Electrical Engineering, Southern University of Science and Technology, Shenzhen 518055, China (e-mail: youcs@sustech.edu.cn).} }
	
	\maketitle
	
	\begin{abstract}
		Enhancing user fairness in advanced multi-user systems like two-layer rate-splitting multiple access (RSMA) is a critical yet challenging task. This letter proposes a novel movable antenna (MA) approach to address this challenge. We formulate a max-min fairness problem, maximizing the minimum user rate, a key metric for fairness, through the joint optimization of the beamforming matrices, user clustering, common rate allocation, and the antenna position vector (APV). To solve this non-convex problem, we develop an efficient two-loop iterative algorithm. The outer-loop leverages the dynamic neighborhood pruning particle swarm optimization method to find a high-quality APV, while the inner-loop optimizes the remaining variables for a given APV. Simulation results validate our approach, demonstrating that the proposed scheme yields significant fairness gains over various benchmark schemes.
	\end{abstract}
	\section{Introduction}
	\IEEEPARstart{R}{ate}-splitting multiple access (RSMA) has emerged as a powerful non-orthogonal transmission framework for sixth-generation (6G) networks, primarily due to its flexibility in managing interference \cite{1}. By splitting messages into common and private streams, it provides a unified structure that contains space-division multiple access (SDMA) and non-orthogonal multiple access (NOMA) as special cases. This adaptability has been shown to yield superior gain \cite{2}.
	
	A key limitation of conventional one-layer RSMA is that its common rate is dictated by the user with the weakest channel, creating a fairness bottleneck in large systems. The two-layer RSMA architecture mitigates this by introducing a hierarchical structure with inter- and intra-cluster common streams, enabling more efficient interference management via two-layer successive interference cancellation (SIC). While this has been shown to improve efficiency \cite{3}, \cite{4}, the ultimate system performance, and by extension, the achievable user fairness, remains fundamentally constrained by the static wireless channel.
	
	Recently, movable antenna (MA) technology has been introduced as a new paradigm to proactively engineer the wireless channel \cite{5}. By dynamically adjusting antenna positions, MAs can create more favorable channel conditions, offering additional degrees of freedom (DoFs) for communication design. This flexibility provides significant performance enhancements over traditional fixed-position antennas (FPAs) in terms of beamforming, spatial multiplexing, and diversity gains \cite{5}--\cite{7}.
	
	The synergy between MAs and RSMA is a promising but nascent research area. Existing works have primarily focused on MA-assisted one-layer RSMA or sum-rate-oriented designs \cite{8}, \cite{9}. The potential of MAs in enhancing user fairness within the two-layer RSMA framework remains largely unexplored. This letter bridges this gap by proposing an MA approach specifically designed to enhance user fairness in two-layer RSMA system. The main contributions are threefold. First, we integrate MAs into a  two-layer RSMA system, alleviating the common rate bottleneck in conventional one-layer RSMA system. Second, we formulate a joint optimization of the beamforming matrices, common rate allocation, user clustering, and the antenna position vector (APV) from a user fairness perspective. Third, to solve this non-smooth and non-convex problem, we develop an efficient two-loop iterative algorithm. The outer-loop optimizes APV using dynamic neighborhood pruning particle swarm optimization (DNPPSO) algorithm, and the inner-loop determines user clustering and resource allocation via channel similarity and successive convex approximation (SCA) techniques. Simulation results demonstrate that our proposed scheme achieves significant fairness gains compared to various benchmark schemes.
	\section{System Model and Problem Formulation}
	\begin{figure}[!t]
		\centering
		\includegraphics[width=0.9\linewidth]{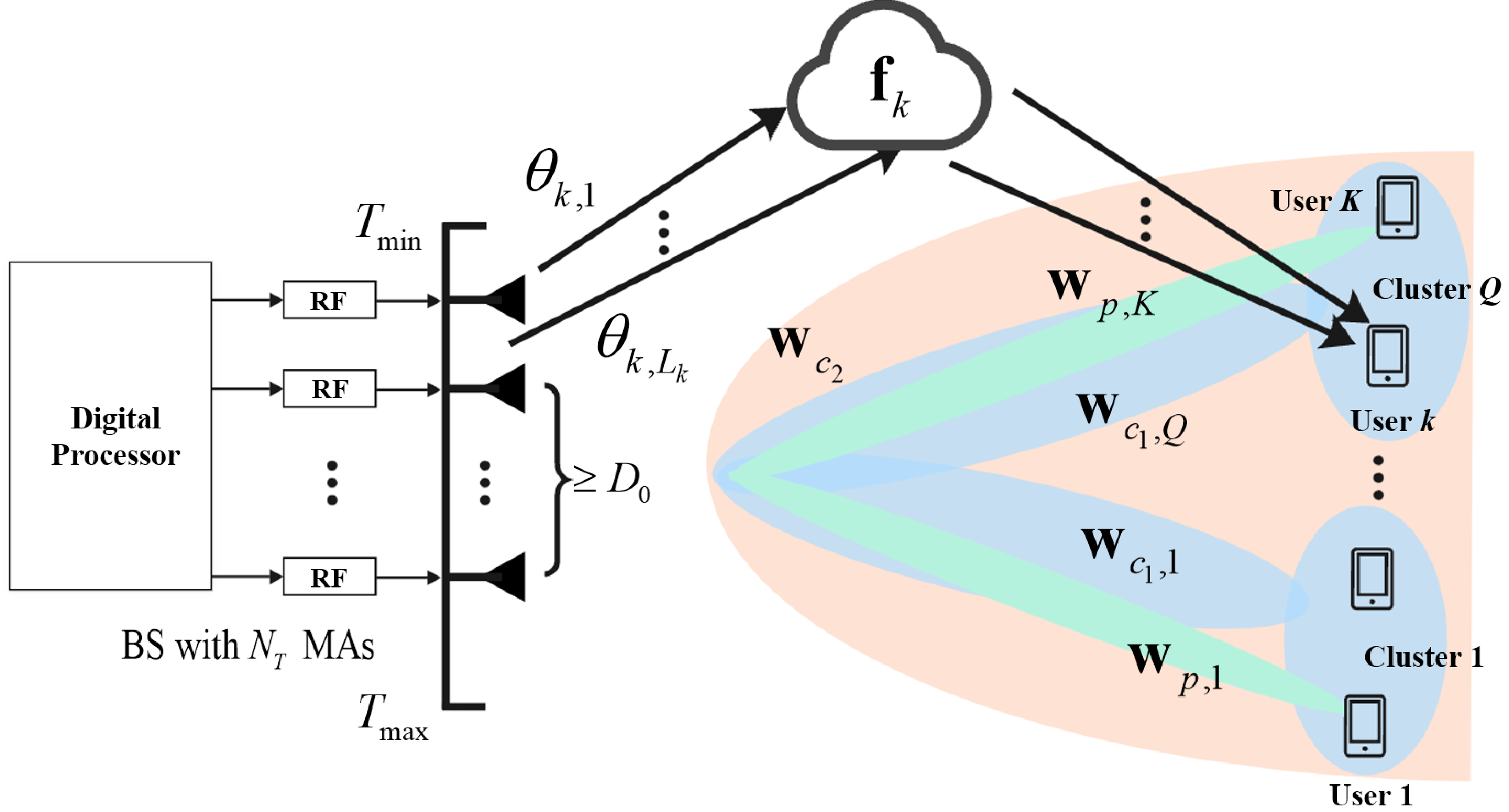}
		\caption{An MA-assisted two-layer RSMA system.}
		\label{fig:1}
	\end{figure}
		As illustrated in Fig. 1, we consider a downlink communication system where a base station (BS), equipped with $N_T$ MAs, serves $K$ single-antenna users via a two-layer RSMA scheme. The sets of indices for the MAs and users are denoted by $\mathcal{M} \triangleq \{1, 2, \ldots, N_T\}$ and $\mathcal{K} \triangleq \{1, 2, \ldots, K\}$, respectively.
		
		Following \cite{7}, we use a one-dimensional (1D) local coordinate system to specify the MA positions. The APV is defined as $\mathbf{t} \triangleq [t_1, t_2, \ldots, t_{N_T}]$, where the coordinate $t_m$ of the $m$-th MA is constrained by:
		\begin{align}
			T_{\min} & \leq t_m \leq T_{\max},  \; \forall m\in\mathcal{M},\label{PositionRegion}  \\
			|t_i - t_j| & \geq D_0, \;  \forall i,j\in\mathcal{M}, \; i\neq j.\label{MADistance}
		\end{align}
		Here, $T_{\text{min}}$ and $T_{\text{max}}$ define the boundaries of linear region for the MA array, and $D_0$ is the minimum spacing that needs to be maintained between each MA to mitigate coupling effects, which implies that at most $N_{T}^{\text{max}}=\left\lfloor \frac{T_{\max} - T_{\min}}{D_0} \right\rfloor+1$ MAs can be accommodated in this region, where \( \left\lfloor \cdot \right\rfloor \) denotes the floor operator.
		
		For the implementation of the two-layer RSMA scheme \cite{1}, the users are partitioned into $Q$ disjoint clusters. Let $\mathcal{Q}\triangleq\{1,2,\ldots,Q\}$ be the set of cluster indices. Each cluster $q \in \mathcal{Q}$ contains a subset of users $\mathcal{K}_q$, such that:
		\begin{equation}
			\begin{aligned}\label{cluster}
				\bigcup_{q\in\mathcal{Q}} \mathcal{K}_q = \mathcal{K},\quad \mathcal{K}_q \cap \mathcal{K}_{q'} = \emptyset,\forall q, q' \in \mathcal{Q}, q \neq q'.
			\end{aligned}
		\end{equation}
		
		The transmitted signal consists of three types of streams: one inter-cluster common stream $s_{c_2}$ decoded by all users; $Q$ intra-cluster common streams $\{s_{c_1,q}\}_{q \in \mathcal{Q}}$, where $s_{c_1,q}$ is decoded by users in cluster $q$; and $K$ private streams $\{s_{p,k}\}_{k \in \mathcal{K}}$, where $s_{p,k}$ is decoded only by user $k$. 
		
		Each stream is transmitted with a dedicated beamforming vector, denoted by $\mathbf{w}_{c_2}$, $\mathbf{w}_{c_1,q}$, and $\mathbf{w}_{p,k}$, respectively. The superposed signal at the BS is  $\mathbf{x} = \mathbf{w}_{c_2}s_{c_2} + \sum_{q \in \mathcal{Q}} \mathbf{w}_{c_1,q}s_{c_1,q} + \sum_{k \in \mathcal{K}} \mathbf{w}_{p,k}s_{p,k}.$ The streams are assumed to have normalized average power, and the beamforming vectors are subject to a total transmit power constraint at the BS:
		\begin{equation}
			\|\mathbf{w}_{c_2}\|^2 + \sum_{q\in\mathcal{Q}}\|\mathbf{w}_{c_1,q}\|^2 + \sum_{k\in\mathcal{K}}\|\mathbf{w}_{p,k}\|^2 \leq P_{\text{max}}. \label{powerconstraint}
		\end{equation}
		
		The signal received at user $k$ is given by
		\begin{equation}
			y_k = \mathbf{h}_k^H(\mathbf{t}) \mathbf{x} + n_k,\label{receivedsignal}
		\end{equation}
		where $n_k\sim \mathcal{CN}(0,{\sigma}^2)$ denotes the additive white Gaussian noise and $\mathbf{h}_k(\mathbf{t})$ is the channel vector. The channel from the BS to user $k$ is characterized by $L_k$ paths \cite{6}, where the $\ell$-th path has an angle of departure (AoD) of $\theta_{k,\ell}$. The field response vector (FRV) of the $m$-th MA for user $k$ is
		\begin{equation}
			\mathbf{a}_k(t_m)=[e^{j\frac{2\pi}{\lambda} t_m \cos{\theta_{k,1}}}, \cdots, e^{j \frac{2\pi}{\lambda} t_m \cos{\theta_{k,L_k}}}]^T,\label{FRV}
		\end{equation}
		where $\lambda$ is the wavelength. The collective field response matrix (FRM) for user $k$ is $\mathbf{A}_k(\mathbf{t}) = [\mathbf{a}_k(t_1), \ldots, \mathbf{a}_k(t_{N_T})] \in \mathbb{C}^{L_k \times N_T}$. Consequently, the channel vector is\footnote{We assume a quasi-static channel model and perfect channel state information (CSI), which can be acquired via compressed sensing~\cite{5}, to establish a theoretical performance benchmark. Practical constraints (e.g., movement time, actuation energy, and imperfect CSI) are left for future work.}
		\begin{equation}
			\mathbf{h}_k(\mathbf{t}) = \mathbf{A}_k^H(\mathbf{t}) \boldsymbol{\mathbf{f}}_k \in\mathbb{C}^{N_T \times 1},\label{channelvecotr}
		\end{equation}
		where \(\mathbf{f}_{k}=[f_{k,1},f_{k,2},\ldots,f_{k,L_{k}}]^{T}\in \mathbb{C}^{L_{k} \times 1}\) is the path-response vector (PRV), with $f_{k,\ell}$ representing the complex gain of the $\ell$-th path for user $k$.
		
		Each user $k \in \mathcal{K}_q$ employs SIC to decode its intended streams. The decoding order is: 1) the inter-cluster common stream $s_{c_2}$, 2) the intra-cluster common stream $s_{c_1,q}$, and 3) the private stream $s_{p,k}$. At each stage, the successfully decoded signal is subtracted from the received signal before decoding the next stream. The signal-to-interference-plus-noise ratios (SINRs) for user $k$ to decode $s_{c_2}$, $s_{c_1,q}$, and $s_{p,k}$ are respectively given by
		\begin{equation}\label{SINR}
			\begin{aligned}[b]
				\gamma_{c_2,k} &= \frac{|\mathbf{h}_k^H(\mathbf{t}) \mathbf{w}_{c_2}|^2}{I_k},\gamma_{c_1,q_k} = \frac{|\mathbf{h}_k^H(\mathbf{t}) \mathbf{w}_{c_1,q}|^2}{I_k - |\mathbf{h}_k^H(\mathbf{t}) \mathbf{w}_{c_1,q}|^2},\\
				\gamma_{p,k} &= \frac{|\mathbf{h}_k^H(\mathbf{t}) \mathbf{w}_{p,k}|^2}{I_k -|\mathbf{h}_k^H(\mathbf{t}) \mathbf{w}_{c_1,q}|^2- |\mathbf{h}_k^H(\mathbf{t}) \mathbf{w}_{p,k}|^2},
			\end{aligned}
		\end{equation}
		where $I_k=\sum_{q\in\mathcal{Q}}\left|\mathbf{h}_k{}^H(\mathbf{t})\mathbf{w}_{c_1,q}\right|^2+\sum_{k\in\mathcal{K}}\left|\mathbf{h}_k{}^H(\mathbf{t})\mathbf{w}_{p,k}\right|^2+\sigma^{2}$ denotes the interference power from all intra-cluster and private streams plus noise power when decoding $s_{c_2}$. The corresponding achievable rates are $R_{c_2,k}=\log_2(1+\gamma_{c_2,k})$, $R_{c_1,q_k}=\log_2(1+\gamma_{c_1,q_k})$, and $R_{p,k}=\log_2(1+\gamma_{p,k})$.
		
		For the common streams to be decodable by all their intended users, their rates are limited by the user with the worst channel condition. That is, the rate of the inter-cluster stream is $R_{c_2} = \min_{k \in \mathcal{K}} R_{c_2,k}$, and the rate of the intra-cluster stream for cluster $q$ is $R_{c_1,q} = \min_{k \in \mathcal{K}_q} R_{c_1,q_k}$. These common rates are then allocated among the respective users. Let $r_{c_2,k}$ and $r_{c_1,q_k}$ be the portions of the common rates allocated to user $k$. The following constraints must hold:
		\begin{equation}
			\label{rateconstraint}
				\begin{aligned}[b]
					\sum_{k\in\mathcal{K}}r_{c_2,k} &\leq R_{c_2}, \quad \sum_{k\in\mathcal{K}_q}r_{c_1,q_k} \leq R_{c_1,q}, \forall q\in\mathcal{Q},\\
					r_{c_2,k} &\geq 0, \quad r_{c_1,q_k} \geq 0, \quad \forall k\in\mathcal{K}_q, \forall q\in\mathcal{Q}.
			\end{aligned}
		\end{equation}		
		The total rate of user $k$ is the sum of its allocated common rates and its private rate, i.e., $R_{{q_k}} = r_{c_2,k} + r_{c_1,q_k} + R_{p,k}$.
		
		The minimum user rate depends on the beamforming matrix $\mathbf{W} \triangleq [\mathbf{w}_{c_2}, \{\mathbf{w}_{c_1,q}\}, \{\mathbf{w}_{p,k}\}]\in \mathbb{C}^{N_T\times (1+Q+K)}$, the user clustering $\{\mathcal{K}_q\}$, the common rate allocation vector $\mathbf{r} \triangleq [\{r_{c_2,k}\},\{r_{c_1,q_k}\}]\in \mathbb{R}^{1\times2K}$, and the APV $\mathbf{t}$. To ensure user fairness, we aim to maximize the minimum user rate by jointly optimizing these variables. The problem is as follows
		\begin{subequations}\label{overallproblem}
			\begin{align}
				\max_{\mathbf{W},\mathbf{t},\mathbf{r},\left\{\mathcal{K}_q\right\}}&\min_{k\in\mathcal{K}}R_{q_{k}}& \label{15a}\\
				\text{s.t.}~~~& \eqref{PositionRegion}, \eqref{MADistance}, \eqref{cluster}, \eqref{powerconstraint}, \eqref{rateconstraint}.\nonumber
			\end{align}
		\end{subequations}
		 Problem \eqref{overallproblem} is challenging to solve due to the non-convex max-min objective function and the intricate coupling among the high-dimensional optimization variables. In the following section, we propose an iterative algorithm to find a high-quality suboptimal solution.
		 \section{Proposed Algorithm for Problem \eqref{overallproblem}}
		 To solve problem \eqref{overallproblem}, we develop a two-loop iterative algorithm that decouples the optimization variables. Specifically, the outer-loop optimizes the APV $\mathbf{t}$ using the DNPPSO algorithm. For each candidate APV generated in the outer-loop, the inner-loop evaluates its fitness value by optimizing the beamforming matrix $\mathbf{W}$, the common rate allocation vector $\mathbf{r}$, and the user clustering $\{\mathcal{K}_q\}$ for the given $\mathbf{t}$.
		 \subsection{Outer-Loop Optimization}
		 In the outer-loop, we optimize the APV $\mathbf{t}$ by addressing the following problem:
		 \begin{subequations}\label{antennaproblem}
		 	\begin{align}
		 		\max_{\mathbf{t}}~&\min_{k\in\mathcal{K}}R_{q_{k}}& \label{16a}\\
		 		\text{s.t.}~&\eqref{PositionRegion},\eqref{MADistance}.\nonumber
		 	\end{align}
		 \end{subequations}
		 Since the objective function \eqref{antennaproblem} is non-differentiable, we employ the DNPPSO algorithm, a gradient-free method. DNPPSO utilizes a swarm of particles, each representing a candidate APV, to explore the solution space. Initially, a population of $P$ particles is created. For the $p$-th particle, its position and velocity vectors are initialized as:
		  \begin{equation}\label{particleinit}
		 	\begin{aligned}
		 		{\mathbf{t}}_p^{(0)}=[{t_{p,1}^{(0)}},{t_{p,2}^{(0)}},...,{t_{p,N_T}^{(0)}}],
		 		{\mathbf{v}}_p^{(0)}=[{v_{p,1}^{(0)}},{v_{p,2}^{(0)}},...,{v_{p,N_T}^{(0)}}], 
		 	\end{aligned}
		 \end{equation}
		 where each element $t_{p,m}^{(0)}$ is randomly chosen from $[T_{\min}, T_{\max}]$ to satisfy \eqref{PositionRegion}.
		 To guide the search, we define a fitness function that incorporates a penalty for constraint violation: \begin{equation}\label{fitness}
		 	\mathcal{F}\left({\mathbf{t}}_p^{(i)}\right) \triangleq R\left({\mathbf{t}}_p^{(i)}\right)-\tau\mathcal{L}\left({\mathbf{t}}_p^{(i)}\right),
		 \end{equation}
		 where $R({\mathbf{t}}_p^{(i)})$ is the minimum user rate obtained from the inner-loop optimization for a given APV ${\mathbf{t}}_p^{(i)}$ at iteration $i$. The term $\mathcal{L}({\mathbf{t}}_p^{(i)})$ counts the number of MAs violating the constraint \eqref{MADistance}, and $\tau$ is a large positive penalty factor. This formulation steers particles towards the feasible region.
		 
		 Each particle updates its velocity and position based on its personal best position, $\mathbf{t}_{p,\text{pbest}}$, and the global best position found so far, $\mathbf{t}_{\text{gbest}}$. The update rules for each iteration are:
		 \begin{align}
		 	{\mathbf{v}}_p^{(i+1)} &= \omega{\mathbf{v}}_p^{(i)} + c_1\tau_1\left({\mathbf{t}}_{p,\text{pbest}}-{\mathbf{t}}_p^{(i)}\right)+ c_2\tau_2\left({\mathbf{t}}_{\text{gbest}}-{\mathbf{t}}_p^{(i)}\right),\label{vupdate}\\
		 	{\mathbf{t}}_p^{(i+1)}&=\mathcal{B}\left({\mathbf{t}}_p^{(i)}+{\mathbf{v}}_p^{(i+1)}\right),\label{positionupdate}
		 \end{align}
		 where $c_1$ and $c_2$ are the personal and global learning factors, respectively, $\tau_1, \tau_2 \sim U(0,1)$ are random parameters, and $\omega$ is an inertia weight that decreases linearly from $\omega_{\text{max}}$ to $\omega_{\text{min}}$ over $I$ iterations. The function $\mathcal{B}(\cdot)$ ensures that the updated positions remain satisfied \eqref{PositionRegion}.
		 A significant computational burden arises from executing the inner-loop optimization for every particle at each iteration. To alleviate this, we incorporate the pruning mechanism from DNPPSO \cite{7}. The rationale is that particles converging near the global best position have a diminished potential for discovering superior solutions. Thus, we define a dynamic neighborhood radius $d_{\mathrm{R}}^{(i)}$ at iteration $i$ and a corresponding neighborhood set $\mathcal{P}(d_{\mathrm{R}}^{(i)})$:
		 \begin{equation}
		 	\bigg\{d_{\mathrm{R}}^{(i)}|~ \sum_{p=1}^{\tilde{P}^{(i)}}\xi\left(\left\|{\mathbf{t}}_{p}^{(i)}-{\mathbf{t}}_{\text{gbest}}\right\|_{2}<d_{\mathrm{R}}^{(i)}\right) \triangleq \tilde{P}^{(i)}-\tilde{P}^{(i+1)}\bigg\},\label{neithborhoodradius}
		 \end{equation}
		 \begin{equation}
		 	\mathcal{P}\left(d_{\mathrm{R}}^{(i)}\right) \triangleq\left\{{\mathbf{t}}_{p}^{(i)}|~\|{\mathbf{t}}_{p}^{(i)}-{\mathbf{t}}_{\text{gbest}}\|_{2}<d_{\mathrm{R}}^{(i)},1\leq p\leq\tilde{P}^{(i)}\right\},\label{neighborhoodset}
		 \end{equation}
		 where $\xi(\cdot)$ is the indicator function and $\tilde{P}^{(i)}$ is the number of active particles at iteration $i$. In each iteration, the neighborhood radius $d_{\mathrm{R}}^{(i)}$ is determined based on the number of additional particles to be pruned, and the particles within this radius are added to the neighborhood set $\mathcal{P}(d_{\mathrm{R}}^{(i)})$, which will skip the inner-loop in the next iteration.
		 \subsection{Inner-Loop Optimization}
		 In the inner-loop, for a given APV $\mathbf{t}_p^{(i)}$, we solve for the user clustering $\{\mathcal{K}_q\}$, transmit beamforming matrix $\mathbf{W}$, and common rate allocation vector $\mathbf{r}$ to evaluate the fitness value in \eqref{fitness}. The subproblem is formulated as
		 \begin{subequations}\label{innerloopproblem}
		 	\begin{align}
		 		\max_{\mathbf{W},\mathbf{r},\left\{\mathcal{K}_q\right\}}&~\min_{k\in\mathcal{K}}R_{q_{k}} \label{26a}\\
		 		\text{s.t.}~~&~\eqref{cluster},\eqref{powerconstraint},\eqref{rateconstraint}. \nonumber
		 	\end{align}
		 \end{subequations}
		 Problem \eqref{innerloopproblem} is challenging due to the coupling between the discrete clustering variables and the continuous beamforming and rate variables. To make it tractable, we decompose it into two subproblems that are solved sequentially: 1) user clustering and 2) joint optimization of beamforming and rate allocation.
		 \subsubsection{User Clustering Optimization}
		 We first determine the user clustering $\{\mathcal{K}_q\}$. Intuitively, to maximize the intra-cluster common rate, users sharing the same beamforming vector $\mathbf{w}_{c_1,q}$ should have highly correlated channel vectors. Thus, we propose a greedy clustering algorithm based on channel similarity. We quantify the similarity between any two users $a$ and $b$ using the cosine similarity metric: $c_{a,b} \triangleq \frac{|\mathbf{h}_a^H\mathbf{h}_b|}{\|\mathbf{h}_a\|\|\mathbf{h}_b\|}$. The algorithm iteratively pairs the two unclustered users with the highest channel similarity. If there is an odd number of users, the last unpaired user is assigned to a separate cluster. The procedure is detailed in Algorithm \ref{algorithm1}.\footnote{To focus on the performance potential of our proposed system, we use fixed-size clustering for its low-complexity. Variable-size clustering is non-trivial due to user-to-cluster similarity design, the risk of reintroducing a common stream bottleneck in large cluster, and the MA-induced dynamic nature; thus, it is left for future work.}
		 \begin{algorithm}[!t]
		 	\caption{Proposed User Clustering Scheme \label{algorithm1}}
		 	\footnotesize
		 	\begin{algorithmic}[1]
		 		\STATE Initialize clusters $\mathcal{K}_q = \emptyset$, $\forall q$, and the set of unclustered users $\mathcal{U} = \mathcal{K}$.
		 		\STATE Calculate matrix $\mathbf{C}$ with its element as $c_{a,b} = \frac{|\mathbf{h}_a^H \mathbf{h}_b|}{\|\mathbf{h}_a\| \|\mathbf{h}_b\|}$ for all $a,b \in \mathcal{U}, a \neq b$.
		 		\STATE Set cluster index $q = 1$.
		 		\REPEAT
		 		\STATE Find $(a^*, b^*) = \arg \max_{a,b \in \mathcal{U}, a \neq b} c_{a,b}$. 
		 		\STATE $\mathcal{K}_q \leftarrow \{a^*, b^*\}$, $\mathcal{U} \leftarrow \mathcal{U} \setminus \{a^*, b^*\}$, $q \leftarrow q + 1$.
		 		\UNTIL{$|\mathcal{U}| < 2$}
		 		\STATE $\mathcal{K}_Q \leftarrow \{u\}$, where $u$ is the remaining element in $\mathcal{U}$.\\
		 		\noindent \textbf{Return:} $\{\mathcal{K}_q\}$.
		 	\end{algorithmic}
		 \end{algorithm}
		 \subsubsection{Joint Beamforming and Rate Allocation Optimization}
		 Then, we proceed to optimize $\mathbf{W}$ and $\mathbf{r}$ by solving
		 \begin{subequations}\label{beamformingandrate}
		 	\begin{align}
		 		\max_{\mathbf{W},\mathbf{r},}&~\min_{k\in\mathcal{K}}R_{q_{k}}\\
		 		\text{s.t.}&~\eqref{powerconstraint},\eqref{rateconstraint}.\nonumber
		 	\end{align}
		 \end{subequations}
		 Problem \eqref{beamformingandrate} is non-convex. To tackle it, we first apply semidefinite relaxation (SDR). We define $\mathbf{W}_{c_2} \triangleq \mathbf{w}_{c_2}\mathbf{w}_{c_2}^H$, $\mathbf{W}_{c_1,q} \triangleq \mathbf{w}_{c_1,q}\mathbf{w}_{c_1,q}^H$, and $\mathbf{W}_{p,k} \triangleq \mathbf{w}_{p,k}\mathbf{w}_{p,k}^H$. Let $\widetilde{\mathbf{W}} \triangleq \{\mathbf{W}_{c_2}, \{\mathbf{W}_{c_1,q}\}, \{\mathbf{W}_{p,k}\} \}$ be the set of these new matrix variables, where $\widetilde{\mathbf{W}}_i$ represents the $i$-th matrix in $\widetilde{\mathbf{W}}$, i.e., $\widetilde{\mathbf{W}}_1 = \mathbf{W}_{c_2}$, $\widetilde{\mathbf{W}}_{1+q} = \mathbf{W}_{c_1,q}$, and $\widetilde{\mathbf{W}}_{1+Q+k} = \mathbf{W}_{p,k}$. The sum of these new matrix variables is defined as $\widetilde{\mathbf{W}}_s = \mathbf{W}_{c_2} + \sum_{q \in \mathcal{Q}} \mathbf{W}_{c_1,q} + \sum_{k \in \mathcal{K}} \mathbf{W}_{p,k}$. By introducing a slack variable $z$ for the minimum rate, problem \eqref{beamformingandrate} is recast as:
		      \begin{subequations}\label{SDR}
		 	\begin{align}
		 		\max_{\widetilde{\mathbf{W}},\mathbf{r},z}~&{z}\\
		 		\label{tracepower}\text{s.t.}~&\mathrm{tr}\left(\widetilde{\mathbf{W}}_s\right) \leq P_{\mathrm{max}},\\
		 		\label{sdrprivate}&r_{c_2,k} + r_{c_1,q_k} +\Gamma_{p,k} + \Lambda_{p,k} \geq z,\forall k\in\mathcal{K},\\
		 		\label{sdrinter}&\Gamma_{c_2,k} + \Lambda_{c_2,k} \geq \sum_{k\in\mathcal{K}} r_{c_2,k},\forall k \in \mathcal{K},\\
		 		\label{sdrintra}&\Gamma_{c_1,q_k} + \Lambda_{c_1,q_k} \geq \sum_{k\in\mathcal{K}_q} r_{c_1,q_k},\forall k \in \mathcal{K}_q,q\in\mathcal{Q},\\
		 		\label{semide}&\widetilde{\mathbf{W}}_i \succeq 0 ,\forall i \in \{1, \ldots, 1+Q+K\},\\
		 		\label{rankone}&\text{rank}\left(\widetilde{\mathbf{W}}_i\right) = 1,\forall i \in \{1, \ldots, 1+Q+K\},\\
		 		&r_{c_2,k} \geq 0, r_{c_1,q_k} \geq 0, \forall k\in\mathcal{K}_q, \forall q\in\mathcal{Q}.\label{biggerthan0}
		 	\end{align}
		 \end{subequations}
		  where the functions in \eqref{sdrprivate}-\eqref{sdrintra} are defined as $\Gamma_{l}\triangleq \log_2(\lambda_{l} + \sigma^2), (l \in \{\{c_2,k\}, \{c_1,q_k\}, \{p,k\}\})$, $\Lambda_{p,k} \triangleq -\log_2(\varphi_{p,k} + \sigma^2)$, $\Lambda_{c_2,k} \triangleq -\Gamma_{c_1,q_k}$, $\Lambda_{c_1,q_k} \triangleq -\Gamma_{p,k}$, and
		 \begin{equation}\label{lambdaandvarphi}
		 	\begin{aligned}[b]
		 		\lambda_{c_2,k} &= \mathbf{h}_k^H \widetilde{\mathbf{W}}_s \mathbf{h}_k,~\lambda_{c_1,q_k} = \mathbf{h}_k^H \left( \widetilde{\mathbf{W}}_s - \widetilde{\mathbf{W}}_1 \right) \mathbf{h}_k\\
		 		\lambda_{p,k} &= \mathbf{h}_k^H \left( \widetilde{\mathbf{W}}_s - \widetilde{\mathbf{W}}_1 - \widetilde{\mathbf{W}}_{1+q} \right) \mathbf{h}_k\\
		 		\varphi_{p,k} &= \mathbf{h}_k^H \left( \widetilde{\mathbf{W}}_s - \widetilde{\mathbf{W}}_1 - \widetilde{\mathbf{W}}_{1+q} - \widetilde{\mathbf{W}}_{1+Q+k} \right) \mathbf{h}_k.
		 	\end{aligned}
		 \end{equation}
		 Problem \eqref{SDR} remains non-convex due to the rank-one constraint \eqref{rankone} and the rate constraints, which have a difference of convex (D.C.) structure. We address the D.C. constraints using the SCA technique. Specifically, the convex terms 
		 	$\Lambda_{c_{2},k}$, $\Lambda_{c_{1},q_{k}}$, and $\Lambda_{p,k}$ are replaced by their 
		 	first-order Taylor lower bounds around the solution obtained at the previous iteration,\footnote{In the first SCA iteration, the common stream beamformers are initialized as zero, while the private stream beamformers are initialized using maximum ratio transmission (MRT) subject to the transmit power constraint.} 
		 	denoted by $(\cdot)^{(j)}$. The lower bound of $\Lambda_{c_2,k}$ is given by
		 	\begin{align}
		 		\Lambda_{c_2,k}^{(j,lb)} &= -\log_2(\lambda_{c_1,q_k}^{(j)} + \sigma^2) - \frac{\lambda_{c_1,q_k} - \lambda_{c_1,q_k}^{(j)}}{(\lambda_{c_1,q_k}^{(j)} + \sigma^2) \ln 2}.\label{firstorder}
		 	\end{align}
		 	Addressing $\Lambda_{c_1,q_k}$ and $\Lambda_{p,k}$ is in a similar manner and thus is omitted for brevity. By substituting these bounds, the problem to be solved at iteration $j+1$ is
		 	\begin{subequations}\label{SCA}
		 		\begin{align}
		 			\max_{\widetilde{\mathbf{W}},\mathbf{r},z}\ &{z}\\
		 			\label{scainter}\text{s.t.}~&\Gamma_{c_2,k} + \Lambda_{c_2,k}^{(j,lb)} \geq \sum_{k\in\mathcal{K}} r_{c_2,k},\forall k \in \mathcal{K},\\
		 			\label{scaintra}&\Gamma_{c_1,q_k} + \Lambda_{c_1,q_k}^{(j,lb)} \geq \sum_{k\in\mathcal{K}_q} r_{c_1,q_k},\forall k \in \mathcal{K}_q,q\in\mathcal{Q},\\
		 			\label{scaprivate}&r_{c_2,k} + r_{c_1,q_k} +\Gamma_{p,k} + \Lambda_{p,k}^{(j,lb)} \geq z,\forall k\in\mathcal{K},\\
		 			& \eqref{tracepower}, \eqref{semide}, \eqref{rankone}, \eqref{biggerthan0}. \nonumber
		 		\end{align}
		 	\end{subequations}
	    	The only remaining non-convexity is the constraint \eqref{rankone}. Besides, according to Theorem 1 in \cite{2}, the optimal solution of problem \eqref{SCA} without \eqref{rankone}, can always satisfy rank-one constraint. As a result, we can drop the constraint \eqref{rankone} without loss of its optimality and solve the resulting semidefinite programming (SDP) problem via the interior-point method.
	    		\subsection{Overall Algorithm}
	    	    The proposed two-loop iterative algorithm is summarized in Algorithm~\ref{Algorithm2}. 
	    		In the outer-loop, the global best position $\mathbf{t}_{\text{gbest}}$ is updated only when a particle finds a position with a higher fitness value, ensuring the objective value is non-decreasing across iterations. Furthermore, the objective value is upper-bounded due to the power constraint~\eqref{powerconstraint}. Thus, the proposed algorithm exhibits a monotonic and bounded behavior. The user clustering has a complexity of $\mathcal{O}(K^2)$. Solving problem~\eqref{SCA} via interior-point method has a complexity of $\mathcal{O}(\log(1/\epsilon)N_T^{4.5})$, where $\epsilon$ is the solution accuracy~\cite{3}. Therefore, the total computational complexity of Algorithm~\ref{Algorithm2} is $\mathcal{O}((K^2+\log(1/\epsilon)N_T^{4.5})\sum_{i=1}^{I}\tilde{P}^{(i)})$ \cite{7}.
	    		\begin{algorithm}[!t]
	    			\caption{Two-Loop Algorithm for Problem \eqref{overallproblem} \label{Algorithm2}}
	    			\footnotesize
	    			\begin{algorithmic}[1]
	    				\STATE Initialize particle velocities $\{{\mathbf{v}}_p^{(0)}\}_{p=1}^{P}$ and positions $\{{\mathbf{t}}_p^{(0)}\}_{p=1}^{P}$.
	    				\STATE Evaluate initial fitness value $\mathcal{F}({\mathbf{t}}_p^{(0)})$ for each particle via \eqref{fitness}.
	    				\STATE Set ${\mathbf{t}}_{p,\text{pbest}} \leftarrow {\mathbf{t}}_{p}^{(0)}, \forall p$, and ${\mathbf{t}}_{\text{gbest}} \leftarrow \arg \max_{p} \{ \mathcal{F} ( {\mathbf{t}}_{p}^{(0)} ) \}$.
	    				
	    				\FOR{$i = 1:I$}  
	    				\STATE Update inertia weight $\omega$ and number of active particles $\tilde{P}^{(i)}$.
	    				\FOR{$p = 1:P$}
	    				\IF{particle ${\mathbf{t}}_p^{(i-1)}$ was not pruned}
	    				\STATE Update ${\mathbf{v}}_p^{(i)}$ and ${\mathbf{t}}_p^{(i)}$ via \eqref{vupdate} and \eqref{positionupdate}.
	    				\STATE Evaluate fitness $\mathcal{F}({\mathbf{t}}_p^{(i)})$ by solving problem \eqref{innerloopproblem}.
	    				\STATE Update ${\mathbf{t}}_{p,\text{pbest}}$ if $\mathcal{F}({\mathbf{t}}_{p}^{(i)}) > \mathcal{F}({\mathbf{t}}_{p,\text{pbest}})$.
	    				\STATE Update ${\mathbf{t}}_{\text{gbest}}$ if $\mathcal{F}({\mathbf{t}}_{p}^{(i)}) > \mathcal{F}({\mathbf{t}}_{\text{gbest}})$.
	    				\ENDIF
	    				\ENDFOR
	    				\STATE Determine neighborhood radius $d_R^{(i)}$ and set $\mathcal{P}(d_{\mathrm{R}}^{(i)})$ via \eqref{neithborhoodradius}, \eqref{neighborhoodset}.
	    				\STATE Mark particles within $\mathcal{P}(d_{\mathrm{R}}^{(i)})$ to be skipped in the next iteration.
	    				\ENDFOR\\
	    				\noindent \textbf{Return:} final APV $\mathbf{t}= \mathbf{t}_\text{gbest}$ and the corresponding $\mathbf{W}$, $\mathbf{r}$, $\{\mathcal{K}_q\}$
	    			\end{algorithmic}
	    		\end{algorithm}
	\section{Simulation Results}
			 \begin{figure*}[!t]  
		\centering
		\begin{subfigure}[b]{0.255\linewidth}
			\centering
			\includegraphics[width=\linewidth]{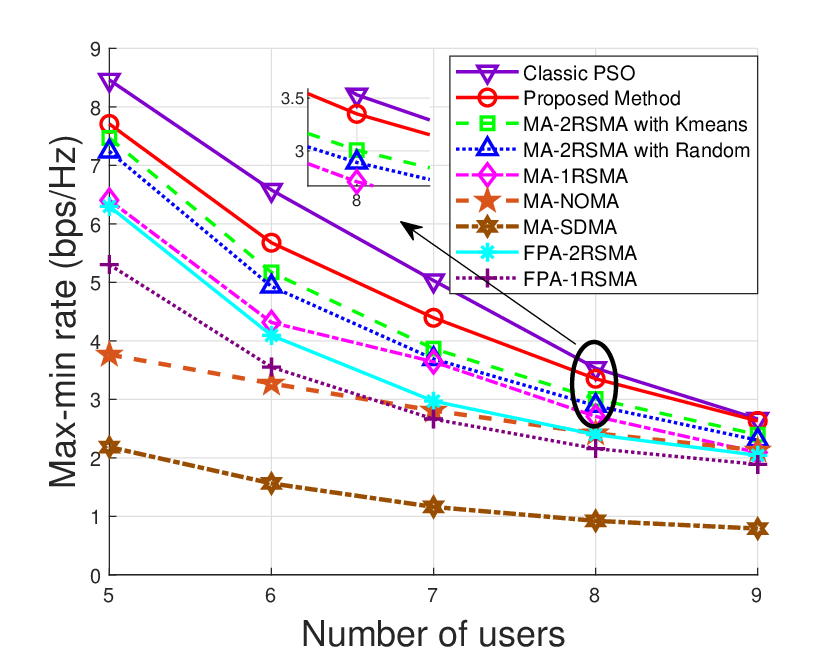}
			\caption{}
			\label{fig:sub1}
		\end{subfigure}
		\hspace{-0.035\linewidth} 
		\begin{subfigure}[b]{0.255\linewidth}
			\centering
			\includegraphics[width=\linewidth]{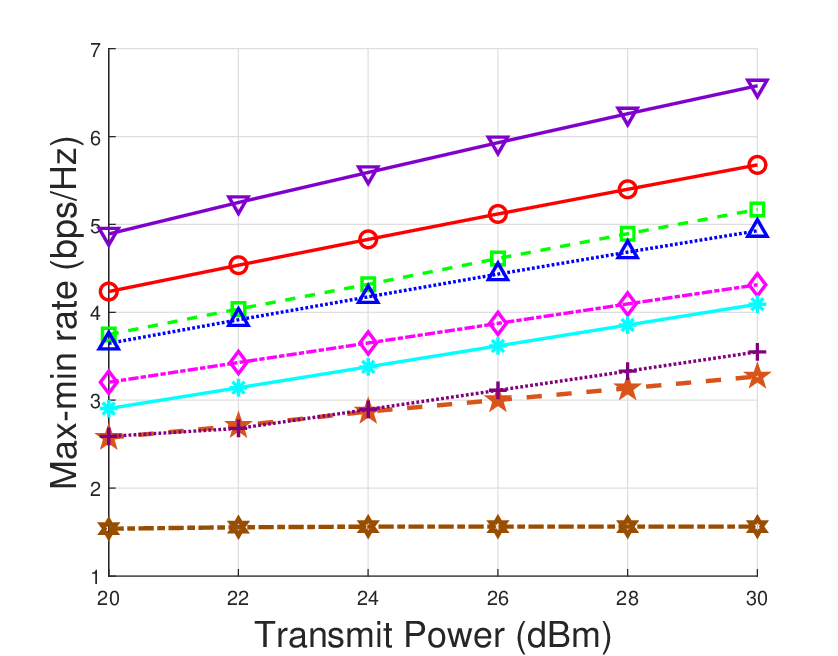}
			\caption{}
			\label{fig:sub2}
		\end{subfigure}
		\hspace{-0.035\linewidth} 
		\begin{subfigure}[b]{0.255\linewidth}
			\centering
			\includegraphics[width=\linewidth]{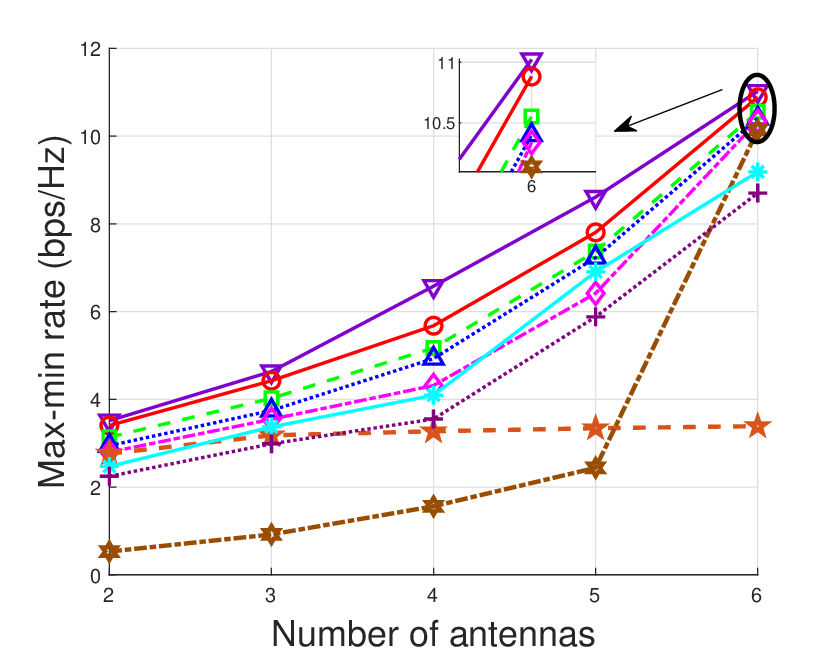}
			\caption{}
			\label{fig:sub3}
		\end{subfigure}
		\hspace{-0.035\linewidth} 
		\begin{subfigure}[b]{0.255\linewidth}
			\centering
			\includegraphics[width=\linewidth]{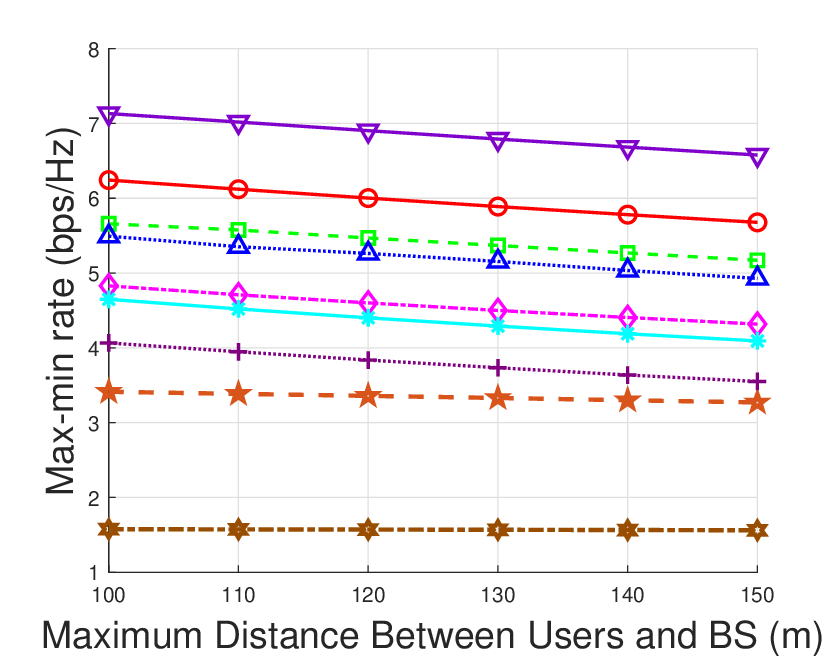}
			\caption{}
			\label{fig:sub4}
		\end{subfigure}
		\caption{Max-min rate versus (a) number of users, with $N_T=4, P_{\text{max}}=30$~dBm; (b) transmit power, with $N_T=4, K=6$; (c) number of antennas, with $K=6, P_{\text{max}}=30$~dBm; and (d) maximum user distance, with $N_T=4, K=6, P_{\text{max}}=30$~dBm. All subfigures share the same legend.}
		\label{fig:four_subfigures}
	\end{figure*}
		In this section, we evaluate the performance of our proposed algorithm. We adopt a geometric channel model with $L_k=L=6$ paths for all users \cite{5}. The elements of the PRV, $f_{k,\ell}$, are modeled as i.i.d. circularly symmetric complex Gaussian (CSCG) random variables, i.e., $f_{k,\ell}\sim\mathcal{CN}\left(0, C_0d_k^{-\alpha}/L_k\right)$, where $C_0=-30$~dB is the channel power gain at a 1~m reference distance, $d_k$ is the distance from the BS to user $k$, and $\alpha=2.8$ is the path loss exponent. Users are uniformly distributed in an angular range of $[0,\pi]$ and $d_k$ uniformly chosen from $[20,150]$~m. The solution accuracy $\epsilon$ for SCA iteration is set to $10^{-3}$. Key system parameters are set as follow: $T_{\min} = 0, T_{\text{max}}=10\lambda$, $D_0 = \lambda/2$, $\lambda=0.1$~m, and $\sigma^{2}=-90$~dBm. The parameters of DNPPSO are set accordingly to commonly used configuration \cite{7}, and the number of particles and iterations are set as $P=50$ and $I=50$, respectively. A linear pruning strategy is used, where the number of active particles decreases to $\beta P$ with $\beta=0.02$, and $c_{1}=c_{2}=1.4$, $\tau=20$, $\omega_{\text{min}}=0.4, \omega_{\text{max}}=0.9$. The results presented below are the average performance over 500 Monte Carlo realizations. In each realization, user locations and channel realizations are randomly generated. To mitigate the impact of random initialization, the outer-loop optimization is executed multiple times with different random seeds for all benchmarks, and the best-performing solution is selected. 
	
	We compare the proposed scheme with several benchmarks:\footnote{We follow a fixed-variable principle in all simulations. For all schemes that required clustering, the same number of clusters and cluster size are used. For all SCA-based optimization, the same iteration settings are adopted. For all scheme employing PSO-based optimization, identical parameters are adopted.}
	\textbf{(1) Classic PSO}: Replaces the DNPPSO method with classic PSO method (i.e., $\mathcal{P}(d_R^{(i)})=\emptyset$).
	\textbf{(2) MA-2RSMA with Kmeans}: Replaces the proposed clustering with the Kmeans algorithm \cite{3}.
	\textbf{(3) MA-2RSMA with Random}: Employs random user clustering.
	\textbf{(4) MA-1RSMA}: A one-layer RSMA baseline where intra-cluster common streams are omitted \cite{9}.
	\textbf{(5) MA-NOMA}: A conventional MA-assisted NOMA scheme \cite{10}.
	\textbf{(6) MA-SDMA}: A SDMA baseline where only private streams are transmitted.
	\textbf{(7) FPA-2RSMA}: The FPA counterpart to our proposed scheme.
	\textbf{(8) FPA-1RSMA}: An FPA-based one-layer RSMA scheme.
	
	Fig. 2(a) illustrates the max-min rate versus the number of users. The proposed method continues to demonstrate competitive performance. The fundamental difference in how MAs benefit these schemes is worth noting. In SDMA, MAs are employed to orthogonalize the channel vectors of different users, whereas in the proposed method, the gain stems from increasing the channel correlation among a subset of users while reducing it for others, coupled with the user clustering.
		
    Fig. 2(b) shows the max-min rate versus BS transmit power. Classic PSO achieves the highest performance, followed by the proposed method, as DNPPSO trades marginal performance for significantly reduced complexity. Despite this trade-off, the proposed method still substantially outperforms other benchmarks due to three factors: MAs' additional spatial DoFs, the two-layer RSMA's superior interference management, and effective channel-aware user clustering.
    
    Fig. 2(c) plots the max-min rate versus the number of transmit antennas. Increasing antennas enhances array and beamforming gains, improving all schemes' performance. The proposed method consistently outperforms benchmarks, achieving target rates with fewer antennas, highlighting its potential for reducing hardware costs.
	
	Fig. 2(d) demonstrates the impact of user distance on the max-min rate. The performance gap between the proposed two-layer RSMA and the one-layer RSMA schemes is particularly pronounced. In one-layer RSMA, the common rate is constrained by the user with the worst channel condition, but our two-layer approach uses intra-cluster streams to manage interference more effectively within smaller user groups.
	\section{Conclusion}
	In this letter, we investigated an MA-assisted two-layer RSMA system to enhance user fairness. We formulated an optimization problem to maximize the minimum user rate through the joint design of the beamforming matrices, common rate allocation, user clustering, and the APV. To solve this non-convex problem, we proposed an efficient two-loop iterative algorithm. The outer-loop leverages the DNPPSO method to find a high-quality APV, while the inner-loop optimizes the remaining variables using a channel similarity-based clustering scheme and an SCA-based resource allocation algorithm. Simulation results validated the effectiveness of our proposed framework, confirming its significant superiority in improving the minimum user rate over several benchmark schemes.

	\end{document}